\documentclass[journal]{IEEEtran}
\usepackage{cite}

\ifCLASSINFOpdf
  \usepackage[pdftex]{graphicx}
  \graphicspath{{../pdf/}{../jpeg/}{../figures/}}
\else
\fi
\ifCLASSOPTIONcompsoc
\usepackage[caption=false, font=normalsize, labelfont=sf, textfont=sf]{subfig}
\else
\usepackage[caption=false, font=footnotesize]{subfig}
\usepackage{booktabs}
\usepackage{multirow}
\usepackage{color}

\usepackage{amsmath}
\usepackage{amsthm}
\usepackage{algorithm}
\usepackage{algorithmic}

\usepackage{array}

\ifCLASSOPTIONcompsoc
 \usepackage[caption=false,font=normalsize,labelfont=sf,textfont=sf]{subfig}
\else
 \usepackage[caption=false,font=footnotesize]{subfig}
\fi
\usepackage{fixltx2e}
\usepackage{dblfloatfix}

\ifCLASSOPTIONcaptionsoff
 \usepackage[nomarkers]{endfloat}
\let\MYoriglatexcaption\caption
\renewcommand{\caption}[2][\relax]{\MYoriglatexcaption[#2]{#2}}
\fi
\usepackage{url}

\begin{document}
%
\title{Applying Translation Symmetry in Electricity Markets to Virtual Power Plant Profit Allocation}
%
%
%

\author{Ruike Lyu, Chuyi Li, Kedi Zheng, Mengshu Shi, Chongqing Kang,~\IEEEmembership{Fellow,~IEEE} and Hongye Guo*

\thanks{R. Lyu, C. Li, M. Shi, C. Kang, and H. Guo are with the State Key Lab of Power System Operation and Control, Dept. of Electrical Eng., Tsinghua Univ., Beijing, China. K. Zheng is with Beijing VPPTech Co. Ltd.,
State Power Investment Corporation (SPIC), Beijing, China. (Corresponding author: H. Guo, Email: hyguo@tsinghua.edu.cn). This work was supported in part by
the National Natural Science Foundation of China under Grant 52307103, and
in part by Postdoctoral Fellowship Program of China Postdoctoral Science
Foundation under Grant GZC20240787.}}

\maketitle

\begin{abstract}
  Virtual power plants (VPPs) are important for coordinating the rapidly growing portfolios of distributed energy resources (DERs) and enabling them to deliver multiple services to higher-level electricity markets.
  However, profit allocation procedures for VPP participants become increasingly difficult to design in an incentive-compatible manner, owing to the increased market power of DERs within each VPP relative to their direct participation in wholesale markets.
  In this paper, we introduce translation symmetry in electricity markets and apply it to VPP aggregation of DERs for market participation to design an incentive-compatible profit allocation method.
  Under the stated assumptions, we prove that this translation symmetry induces an inductive property: once incentive compatibility holds at an upper level, it propagates to the internal settlements between the VPP and its constituent DERs, thereby supporting incentive compatibility throughout the hierarchy.
  We further show that service prices are invariant across levels, which helps preserve competitive conditions and enables transparent value assessment.
  Theoretical analysis and case studies illustrate how this translation-symmetry-based approach can enable incentive-compatible profit allocation when aggregating DERs to provide multiple services.
\end{abstract}


\ifCLASSOPTIONpeerreview
  \begin{center} \bfseries EDICS Category: 3-BBND \end{center}
\fi
%
\IEEEpeerreviewmaketitle





\section{Introduction}
%
%
%
%
\IEEEPARstart{T}{he} rapid proliferation of distributed energy resources (DERs), such as electric vehicles (EVs) and customer-side energy storage, is reshaping power system operations~\cite{yi_aggregate_2021}.
While small in individual size, their vast number makes direct monitoring and control by grid operators difficult~\cite{lyu_co-optimizing_2023, lyu2023lstn}, and also makes their direct participation in wholesale electricity markets computationally infeasible for centralized market operation.
Therefore, aggregation approaches, such as Virtual Power Plants (VPPs), are widely used to bundle these resources and present them as a single, controllable entity akin to traditional power plants~\cite{liu_integrating_2025}.
However, the successful integration of DERs into electricity markets through VPPs requires addressing key challenges related to market and incentive design.

\begin{figure}[!t]
  \centering
  \includegraphics[width=3.0in]{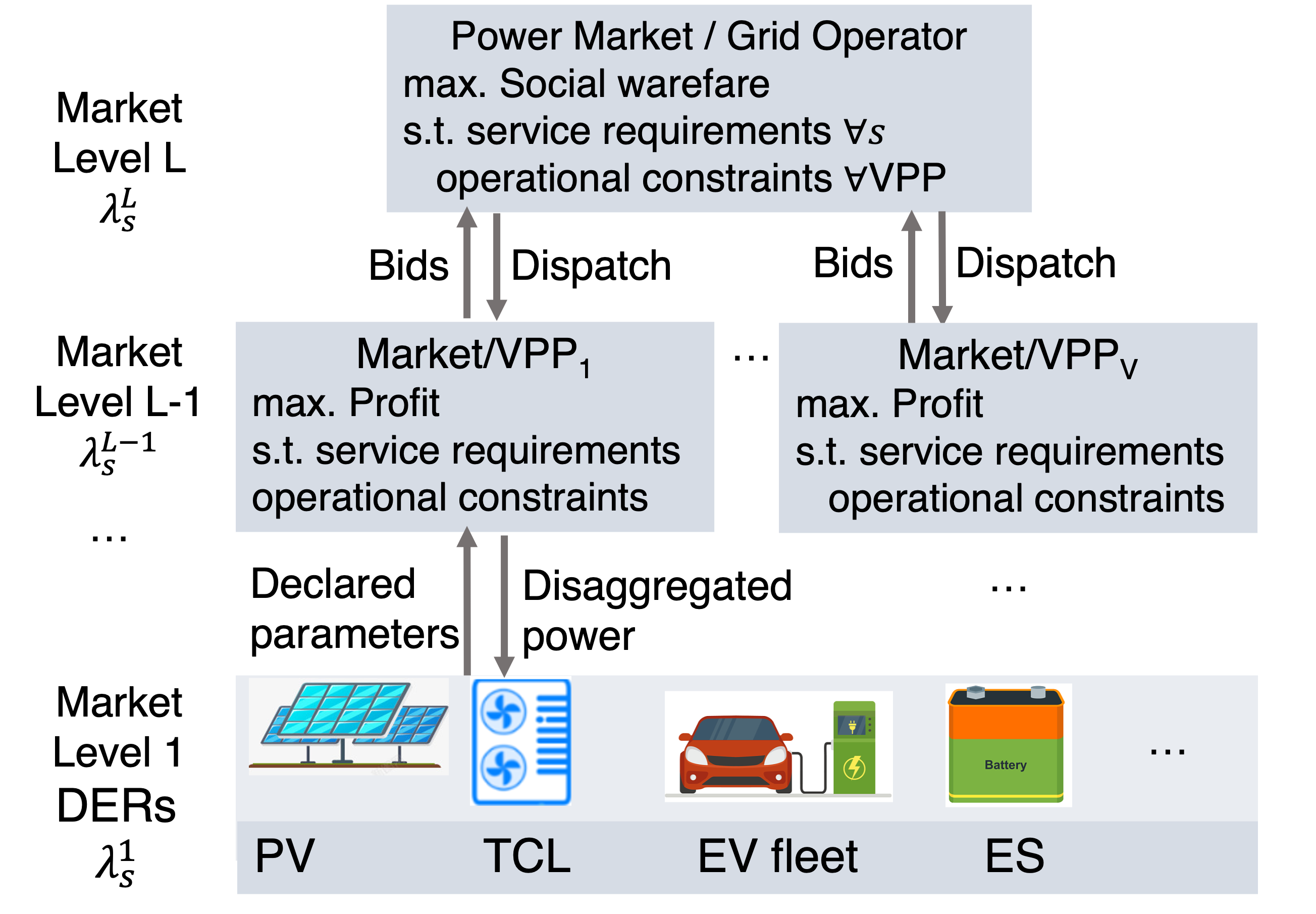}
\caption{Translation symmetry in electricity markets: The market framework employs VPP--DER interaction models with identical structural forms across all hierarchical levels. Under hierarchical translation transformations (adding or removing one level), the model form remains unchanged. This symmetry preserves market properties such as incentive compatibility and invariance of service prices ($\lambda^{l}_s$) throughout the hierarchy.}
  \label{fig_framework}
\end{figure}

The diverse operational flexibility of DERs enables the provision of multiple services to power systems, including energy trading, frequency regulation, reserve services, and ramping capability~\cite{liu_integrating_2025}.
A single resource or combination of different resources can participate in multiple service markets simultaneously, creating opportunities for value maximization through joint participation.
To fully exploit these opportunities, unified VPP operation frameworks are needed that describe different service types using consistent mathematical forms and enable joint optimization across multiple services~\cite{chen_real-time_2024}.
Such VPP frameworks are needed not only for maximizing collective value while respecting operational constraints, but also for implementing a fair profit allocation mechanism that reasonably reflects the diverse contributions of individual resources to the various services provided, thus incentivizing their participation~\cite{zhou_incentive-compatible_2023}.

A central challenge for VPPs is achieving incentive compatibility (IC) for their internal participants. This challenge becomes more pronounced in a hierarchical aggregation framework. By design, aggregation reduces the number of participants in each regional market (VPP), granting individual DERs greater market power than they would have in a flat wholesale market. DER owners can then leverage this power by strategically misreporting private information (e.g., operational costs, capabilities) to maximize their profits under the VPP's allocation rules. The problem is further compounded because a DER's contribution to the VPP's profit from providing multiple services is complex, depending on both the VPP's internal operational decisions and the clearing outcomes of higher-level markets~\cite{robu_rewarding_2016}.

Existing research has widely explored VPP profit allocation, primarily through cooperative game theory approaches such as the nucleolus and Shapley values~\cite{dabbagh2015risk, li2018robust, fang2020improved, DGJS2020S2032, DCYQ202212005}. Refinements include continuous extensions of the Aumann-Shapley procedure to address computational challenges~\cite{li2018robust} and multi-factor methods that account for risk and profit contributions~\cite{DGJS2020S2032, DCYQ202212005}. Other works have used non-cooperative game theory to analyze outcomes among participants who strategically offer parameters~\cite{chao2020profit, singh2021game}. However, despite this progress, current literature (which mainly targets energy markets~\cite{baeyens_coalitional_2013, rahmani-dabbagh_profit_2016}) has devoted limited attention to the specific problem of designing mechanisms that ensure incentive compatibility when DERs strategically misreport their private parameters to the VPP, especially in the context of providing different services simultaneously.

This paper focuses on preserving incentive compatibility when VPPs aggregate DERs to supply multiple services, a setting in which DERs can exert greater market power than in wholesale markets.
To address this issue, we introduce translation symmetry in electricity markets for DER aggregation, in which VPP--DER interaction models maintain identical structural forms across hierarchical levels.
This translation symmetry has two implications.
First, it helps preserve competitive conditions: if a state-level or national-level market is sufficiently competitive, the same conditions can be retained at lower levels, thereby supporting incentive compatibility throughout the hierarchy.
Second, it supports value assessment: resources can infer their system value from transparent price signals, which can guide investment decisions and reduce distortions caused by persistent discrepancies between market payments and underlying resource value.

\section{Hierarchical Market Framework}\label{sec_market}

To illustrate our core concept, we first present a standardized market model for VPPs aggregating distributed resources to participate in higher-level markets.
At the upper level, VPPs submit bids and settle with the wholesale market, while inside each VPP, operation optimization and DER profit allocation are modeled with an analogous clearing-and-settlement structure, ensuring a consistent formulation across layers.
These markets may include energy trading, frequency regulation, reserve services, ramping capability, and other ancillary services.
We refer to these different market types collectively as services, denoted by the index $s \in \mathcal{S}$, where $\mathcal{S}$ represents the set of services.
The market employs a marginal pricing mechanism, where the dual variables of service demand constraints in the VPP--DER interaction problem determine service prices.
Let $\mathcal{V}$ denote the set of VPPs, indexed by $v \in \mathcal{V}$.
Each VPP aggregates distributed energy resources, which we refer to as units, indexed by $u \in \mathcal{U}_v$.
Although the framework can be extended to support multi-level nested hierarchical structures, we first formulate the model using a three-level structure (system level, VPP level, and unit level) for clarity of presentation.
The generalized multi-level formulation will be presented in the next section.

\subsection{The VPP--DER interaction problem} 

The VPP--DER interaction problem determines the optimal allocation of services among participating VPPs.
For VPP $v$ and service $s$, let $x_{v, s}$ denote the service bid quantity, and let $y_v$ represent the internal decision variables of VPP $v$ that are not directly observable to the market operator.
Although we use the notation $v$ for clarity, the same formulation applies to conventional generators that bid directly into the market; in that case, the participant has no subordinate resources but can still be represented by the same bid variables.
The cost for VPP $v$ to provide service $s$ is denoted by $c_{v, s}$.
The service demand, representing the total quantity of service $s$ required by the upper-level market, is denoted by $d_s$.
The VPP--DER interaction problem is formulated as the following optimization problem:
\begin{equation}
\min_{x, y} \quad \sum_{s \in \mathcal{S}} \sum_{v \in \mathcal{V}} c_{v, s} x_{v, s}
\label{eq:market_clearing_problem}
\end{equation}
Subject to the service demand constraints:
\begin{equation}
\sum_{v \in \mathcal{V}} x_{v, s} \ge d_{s} : \lambda_{s}, \quad \forall s \in \mathcal{S}
\label{eq:service_demand}
\end{equation}
where $\lambda_s$ is the dual variable associated with the service demand constraint for service $s$, representing the marginal price of service $s$ in the market.
The optimization problem is also subject to public constraints (\ref{eq:public_constraints}) and private constraints (\ref{eq:technical_constraints}) that involve all VPPs:
\begin{equation}
  f(x_{\mathcal{V}, \mathcal{S}}, y) \le 0
  \label{eq:public_constraints}
\end{equation}
\begin{equation}\label{eq:technical_constraints}
g_v\left(x_{v, \mathcal{S}}, y_v\right) \leq 0, \quad \forall v \in \mathcal{V}
\end{equation}
where $f(x_{\mathcal{V}, \mathcal{S}}, y)$ can include system-wide constraints such as transmission line thermal limits, $y$ denotes the decision variables of the system that are observable only to the market operator, and $g_v(x_{v, \mathcal{S}}, y_v)$ can include private constraints such as ramp rate constraints.

\subsection{Settlement rule and optimal bidding of VPPs}

Under the marginal pricing mechanism, the settlement rule for each VPP $v$ is defined based on the market prices and the allocated service quantities $\sum_{s \in \mathcal{S}} \lambda_{s} x_{v, s}$.
The VPP aims to maximize its profit, which is formulated as:
\begin{equation}
\max \quad \sum_{s \in \mathcal{S}} \lambda_{s} x_{v, s}-\sum_{s \in \mathcal{S}} \sum_{u \in \mathcal{U}_v} c_{u, s} x_{u, s}
\label{eq:vpp_optimal_bidding}
\end{equation}
subject to the optimality conditions of the VPP--DER interaction problem (\ref{eq:market_clearing_problem}) and the service requirement constraints:
\begin{equation}
\sum_{u \in \mathcal{U}_v} x_{u, s} \ge x_{v, s} : \lambda_{v, s}, \quad \forall s \in \mathcal{S}
\label{eq:service_requirement}
\end{equation}
where $\lambda_{v, s}$ is the dual variable associated with the service requirement constraint, representing the internal price of service $s$ for VPP $v$, and $x_{u, s}$ denotes the service quantity allocated to unit $u$ for service $s$.
The problem is also subject to VPP's public constraints (\ref{eq:public_constraints_vpp}) and unit's private constraints (\ref{eq:private_constraints_unit}):
\begin{equation}
  f_v\left(x_{\mathcal{U}_v, \mathcal{S}}, y_v\right) \leq 0
  \label{eq:public_constraints_vpp}
\end{equation}
\begin{equation}
g_u\left(x_{u, \mathcal{S}}, y_u\right) \leq 0, \quad \forall u \in \mathcal{U}_v
\label{eq:private_constraints_unit}
\end{equation}
where $f_v(x_{\mathcal{U}_v, \mathcal{S}}, y_v)$ represents VPP-level public constraints, such as distribution network constraints (e.g., power flow limits, voltage limits) and operational constraints (e.g., communication and control system limitations).
The variable $y_v$ denotes VPP $v$'s internal decision variables that are not directly observable to units but are necessary for enforcing the public constraints.

\subsection{Profit allocation and unit offers}

Similar to the settlement rule at the VPP level, the profit allocation rule for each unit $u$ belonging to VPP $v$ is defined based on the internal prices and service allocations $\sum_{s \in \mathcal{S}} \lambda_{v, s} x_{u, s}$.
The unit aims to maximize its profit, which is formulated as:
\begin{equation}
\max \quad \sum_{s \in \mathcal{S}} \lambda_{v, s} x_{u, s} - \sum_{s \in \mathcal{S}} c_{u, s} x_{u, s}
\label{eq:unit_offer}
\end{equation}
subject to the optimality conditions of the VPP optimal bidding problem (\ref{eq:vpp_optimal_bidding}) and the unit's private constraints.

\section{Translation Symmetry in Electricity Markets}\label{sec_unified_formulation}

We now provide a more abstract mathematical description of the hierarchical market structure and introduce the concept of translation symmetry in electricity markets.
Translation symmetry is characterized by VPP--DER interaction models that maintain identical structural forms across all hierarchical levels, i.e., under hierarchical translation transformations (adding or removing one level), the mathematical form of the model remains unchanged.
Consider a market structure with multiple hierarchical levels, indexed by $l \in \mathcal{L}$, where $\mathcal{L}$ represents the set of market levels and $L = |\mathcal{L}|$ denotes the total number of levels. Each VPP at level $l$ can also be viewed as a market, and its VPP--DER interaction problem determines how VPPs at level $l-1$ should be coordinated at level $l$ to maximize their profits in level $l + 1$, formulated as:
\begin{equation}
\max \quad \sum_{s \in \mathcal{S}} \lambda_s^{l+1} x_{s}^{l} - \sum_{s \in \mathcal{S}} \sum_{v \in \mathcal{V}^{l-1}} c_{v, s}^{l - 1} x_{v, s}^{l - 1}
\label{eq:l_optimal_operation_strategy_vpp}
\end{equation}
subject to the optimality conditions of the VPP--DER interaction problem at level $l + 1$, the service requirement constraints (\ref{eq:l_service_requirement}) and the public constraints at level $l$ and the private constraints at level $l-1$ (\ref{eq:l_public_constraints} - \ref{eq:l_private_constraints}):
\begin{equation}
  \sum_{v \in \mathcal{V}^{l-1}} x_{v, s}^{l-1} \ge x_{s}^{l} : \lambda_{s}^{l}, \quad \forall s \in \mathcal{S}
  \label{eq:l_service_requirement}
\end{equation}
\begin{equation}
  f^{l}\left(x_{\mathcal{V}, \mathcal{S}}^{l-1}, y_{\mathcal{V}}^{l-1}\right) \leq 0
\label{eq:l_public_constraints}
\end{equation}
\begin{equation}
  g_v^{l - 1}\left(x_{v, \mathcal{S}}^{l - 1}, y_{v}^{l - 1}\right) \leq 0, \quad \forall v \in \mathcal{V}^{l-1}
\label{eq:l_private_constraints}
\end{equation}
where $x_{\mathcal{V}, \mathcal{S}}^{l-1}=\left(x_{v, \mathcal{S}}^{l-1}\right)_{v \in \mathcal{V}^{l-1}}$ and $y_{\mathcal{V}}^{l-1}=\left(y_{v}^{l-1}\right)_{v \in \mathcal{V}^{l-1}}$ represent the service bids and internal variables for VPPs at level $l-1$, respectively.
When $l=L$, $\lambda_s^{L+1}=0$, $x_{s}^{L}=d_s$.
Due to the translation symmetry property, the optimal bidding or offering problem at any lower level $l-1$ follows the same mathematical structure as the VPP--DER interaction problem (optimal operation strategy of VPPs) at level $l$ described above.
$v^{l-1}$ may represent either individual physical resources (such as DERs) or lower-level VPPs that aggregate multiple resources. 
We make the following assumptions:

\textbf{Assumption 1 (Feasible region aggregation):}
For any feasible solution at level $l$, i.e., $g^{l}_v\left(x_{v, \mathcal{S}}^{l}, y_{v}^{l}\right) \leq 0$, there exists a corresponding feasible solution at level $l - 1$, i.e., $g^{l-1}_v\left(x_{v, \mathcal{S}}^{l-1}, y_{v}^{l-1}\right) \leq 0, \forall v \in \mathcal{V}^{l-1}$, and vice versa:
\begin{equation}
\forall x_{\mathcal{V}, \mathcal{S}}^{l-1}, \left(
\begin{array}{l}
    \exists y_{V}^{l},\, g^{l}_v\left(\sum_{v \in \mathcal{V}^{l-1}} x_{v, \mathcal{S}}^{l-1}, y_{V}^{l}\right) \leq 0 \\
    \iff \exists y_{V}^{l-1},\, g^{l-1}\bigg(x_{v, \mathcal{S}}^{l-1}, y_{v}^{l-1}\bigg) \leq 0, \forall v \in \mathcal{V}^{l-1}
\end{array}
\right)
\label{eq:assumption_1_equiv_enhanced}
\end{equation}

This assumption ensures that the aggregation of resources from level $l$ to level $l+1$ preserves feasibility, which is the aim of feasible region aggregation studies~\cite{lyu_data-driven_2025}.
Assumption 1 also implies that any resource $v$ can represent its cost through the bidding structure of the proposed mechanism.
Under this representation, resources can strategically adjust bid costs, while bid structure (i.e., volume segments) is fixed and truthful.
Actual cost characteristics can be approximated by a sufficient number of volume-price pairs~\cite{zhou_incentive-compatible_2023}.
Mathematically, this approximation is equivalent to representing different volume-price pairs as different resources $v$, and we use this notation throughout for simplicity.

\textbf{Assumption 2 (Cost function aggregation):}
For any feasible solution at level $l$, i.e., $g_v\left(x_{v, \mathcal{S}}^{l}, y_{v}^{l}\right) \leq 0, \forall v \in \mathcal{V}^{l}$, the cost function at level $l+1$ is no greater than the sum of costs at level $l$:
\begin{equation}
c^{l+1}(\sum_{v \in \mathcal{V}^{l}} x_v^l) \leq \sum_{v \in \mathcal{V}^{l}} c^l\left(x_v^l\right)
\label{eq:assumption_2}
\end{equation}

This assumption reflects the standard cost aggregation approach, where the upper-level cost function is defined as the minimum total cost achievable by optimally allocating the aggregate quantity among lower-level resources.

\textbf{Assumption 3 (Complete competition at the top level):}
The bid of any individual resource does not change the marginal price $\lambda_s^L$ at the top level $L$ for any service $s \in \mathcal{S}$.

This assumption ensures that individual VPPs are price-takers at the top level, which is more likely to be true than at a lower level. Under this translation-symmetric structure with the above assumptions, we propose the following propositions.
Unless otherwise stated, the propositions in this section are interpreted under Assumptions 1-3.

\textbf{Proposition 1 (Incentive compatibility (IC) at top level):}
IC holds at the top level $L$.
This means that resources at level $L-1$ have incentives to truthfully report their cost parameters.

\textbf{Proposition 2 (Inductive property of IC):}
If IC holds at level $l$, then IC holds at level $l-1$.

\textbf{Proposition 3 (Invariance of prices):}
The service prices are invariant across all market levels, i.e.,
\begin{equation}
\lambda_s^l = \lambda_s^{l+1}, \quad \forall s \in \mathcal{S}, \forall l \in \mathcal{L}
\label{eq:proposition_3}
\end{equation}

In the appendix, we prove Propositions 1-3 for a simplified case where the constraints are only quantity-limit constraints, which serves as a simple example to illustrate the properties of the translation-symmetry-based mechanism.
Extending these proofs to settings with more general network and operational constraints remains future work.

\section{Case Study}\label{sec_case_study}

We demonstrate the effects of the proposed profit allocation mechanism through a case study involving a small-scale VPP (composed of an ES, an EV fleet, and a TCL) that aggregates these resources to participate in the NYISO energy, frequency regulation, and reserve ancillary service markets.
This case study considers a two-level hierarchical structure. At the upper level, the VPP acts as a price-taker within an assumed competitive ISO market. At the lower level, internal resources submit their parameters to the VPP.
The VPP then compensates these resources based on the shadow prices derived from the relevant constraints in its optimal operation model, as defined by our proposed mechanism.
We analyze and compare the resulting profit allocations under different bidding strategies and allocation methods.

\begin{figure}[!t]
  \centering
  \includegraphics[width=2.5in]{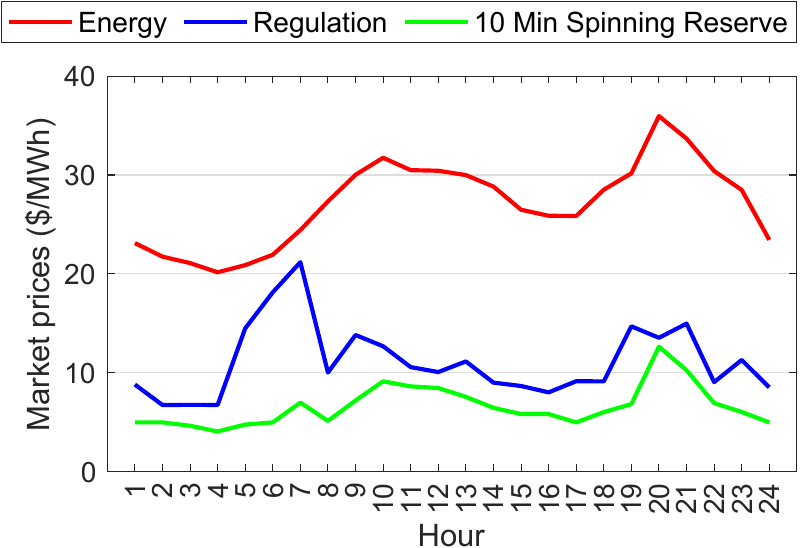}
\caption{Hourly prices ($\lambda^{L}_s$) from NYISO day-ahead market on April 13, 2024, used as top-level market prices.}
  \label{fig_price}
\end{figure}

The ES power/energy capacity is 0.5 MW/1 MWh, the EV fleet consists of 30 bidirectional 7.68 kW EVs, and the TCL has a rated power of 1 MW. The specific parameters are borrowed from \cite{chen_scheduling_2021, lyu_co-optimizing_2023}. We use the day-ahead market prices of NYISO on April 13, 2024 (Fig.~\ref{fig_price}) as the top-level market prices. The specific operation model of the VPP is described in our previous work~\cite{liu_integrating_2025}.

\begin{figure}[!t]
  \centering
  \includegraphics[width=2.5in]{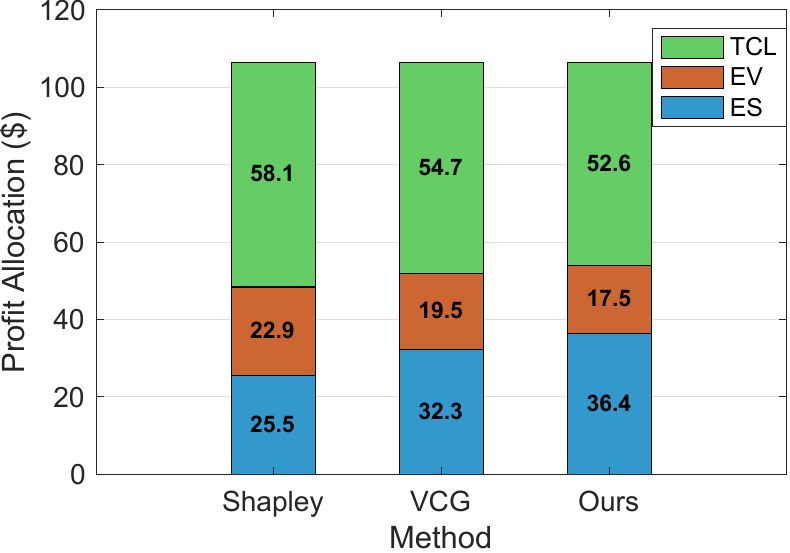}
\caption{Profit allocation results of compared methods.
All three methods yield similar allocations, indicating that the proposed method provides contribution estimates comparable to those of the benchmark methods.}
  \label{fig_profit_allocation}
\end{figure}

Fig. \ref{fig_profit_allocation} presents the AS profit obtained by each resource within the VPP for participating in the ancillary service market under different profit allocation mechanisms. For ease of presentation, the EV fleet is treated as a whole for calculating the benefits. We compared the results using the Shapley value of each resource's contribution to the VPP AS profit~\cite{li2018robust}, the Vickrey-Clarke-Groves (VCG) mechanism~\cite{fang_efficient_2022} (using the change in VPP AS profit before and after the resource's introduction as its contribution), and the proposed method.

The numerical results show that the profit allocations from the Shapley value and VCG methods are close to those from the proposed method, indicating comparable contribution estimates across methods.
Another advantage of the proposed method is that, after VPP bid optimization, it directly yields internal shadow prices for different ancillary services.
In this case study, these shadow prices match the corresponding system-level service values (Fig. \ref{fig_price}), which simplifies the allocation calculation and clarifies its components (the sum of each service contribution multiplied by its corresponding price).

\begin{figure}[!t]
  \centering
  \includegraphics[width=2.5in]{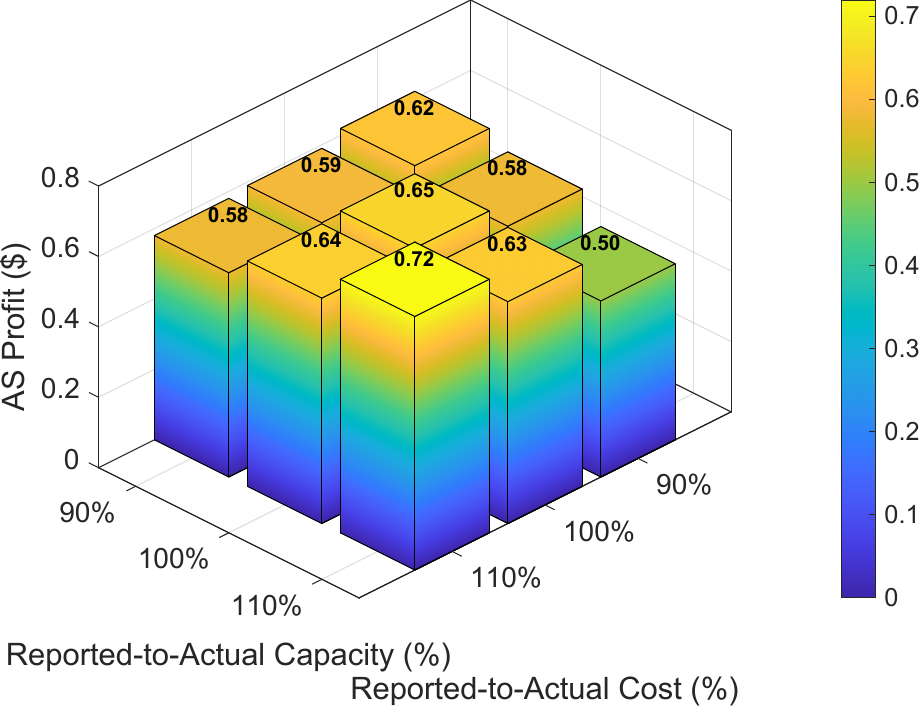}
\caption{VPP profit allocated to EV1 under strategic misreporting scenarios.
Reporting a 10\% increase or decrease in cost or capacity parameters does not yield additional profit, indicating incentive compatibility.
However, simultaneously reporting 110\% of both cost and capacity parameters does increase profit, highlighting the need for performance-based allocation mechanisms.}
  \label{fig_AS_profit_EV1_3D}
\end{figure}

Fig. \ref{fig_AS_profit_EV1_3D} shows the AS profit allocated to a VPP member (EV1) under strategic misreporting of cost and power-capacity parameters.
In this case study, reporting a 10\% increase or decrease from the true parameter values does not increase profit, which is consistent with incentive compatibility under these deviations.
However, simultaneously reporting 110\% of both cost and capacity does increase profit, indicating a limitation that should be addressed through performance-based allocation.
This profitable deviation involves joint misreporting of multiple parameters and highlights a boundary of the current mechanism design.

\section{Conclusion}\label{sec_conclusion}

This paper addresses incentive compatibility in hierarchical market structures for distributed resource aggregation, where greater individual market power at lower levels can weaken truthful reporting incentives.
We introduce translation symmetry in electricity markets, characterized by VPP--DER interaction models that maintain identical structural forms across hierarchical levels.
Under the stated assumptions, the proposed mechanism has two key properties: an inductive property of incentive compatibility from upper levels to lower levels (Proposition 2) and price invariance across market levels (Proposition 3).
Together, these properties help preserve competitive conditions and enable consistent value assessment across levels, provided that sufficient competition exists at the top level.

The propositions presented in this paper are proven for a special case and demonstrated in a toy model case study, which illustrates the main properties of this translation-symmetry-based approach.
Future work will extend the analysis to settings with general network constraints, imperfect top-level competition, and broader empirical tests across multiple market days and resource portfolios.




\appendices
\section{Proof of Proposition 1}\label{app_proof_of_proposition_1}

\begin{proof}
Here we provide the proof of Proposition 1 for level $l=L$ in a special case where the public constraints are empty (e.g., no power flow constraints) and private constraints are only the quantity limit constraints, i.e., $g^{l-1}_v\left(x_{v, \mathcal{S}}^{l - 1}, y_{v}^{l - 1}\right) \leq 0$ takes the form of:
\begin{equation}
  - x_{v, s}^{l - 1} \leq 0 : \underline{\mu}_{v, s}^{l - 1}, \quad x_{v, s}^{l - 1} \leq \overline{x}_{v, s}^{l - 1} : \overline{\mu}_{v, s}^{l - 1}, \quad \forall v, s
\end{equation}
where $\overline{x}_{v, s}^{l - 1}$ is the upper bound of $x_{v, s}^{l - 1}$ and $\underline{\mu}_{v, s}^{l - 1}/\overline{\mu}_{v, s}^{l - 1}$ is the dual variable for the lower/upper bound of $x_{v, s}^{l - 1}$.

Assume the cost function is non-negative. At the optimal point, we have $\sum_v x_{v, s}^{l-1} = d_s$, which turns the service requirement constraint into an equality.
Then the optimality conditions (KKT conditions) of the optimal operation strategy of the VPP at level $l$ are stationary conditions (\ref{eq:l_stationary_condition_vpp}), dual variable non-negativity conditions (\ref{eq:dual_variable_non_negativity_condition}), and complementary slackness conditions (\ref{eq:complementary_slackness_condition}).
\begin{equation}\label{eq:l_stationary_condition_vpp}
  \frac{\partial L}{\partial x_{v, s}^{l - 1}} = c_{v, s}^{l - 1} - \lambda_s^{l} - \underline{\mu}_{v, s}^{l - 1} + \overline{\mu}_{v, s}^{l - 1} = 0, \quad \forall v, s
\end{equation}\vspace{-2ex}
\begin{equation}\label{eq:dual_variable_non_negativity_condition}
  (\overline{\mu}_{v, s}^{l - 1} \ge 0, \underline{\mu}_{v, s}^{l - 1} \ge 0), \quad \forall v, s
\end{equation}\vspace{-2ex}
\begin{equation}\label{eq:complementary_slackness_condition}
  \overline{\mu}_{v, s}^{l - 1} \left(x_{v, s}^{l - 1} - \overline{x}_{v, s}^{l - 1}\right) = 0, \underline{\mu}_{v, s}^{l - 1} \left(x_{v, s}^{l - 1}\right) = 0, \quad \forall v, s
\end{equation}

Under Assumption 3, $\lambda_s^L$ is independent of any individual resource $v$'s bid.
From (\ref{eq:l_stationary_condition_vpp}), (\ref{eq:complementary_slackness_condition}), and (\ref{eq:dual_variable_non_negativity_condition}): if $c_{v, s}^{l - 1} < \lambda_s^{l}$, then $x_{v, s}^{l - 1} = \overline{x}_{v, s}^{l - 1}$; if $c_{v, s}^{l - 1} > \lambda_s^{l}$, then $x_{v, s}^{l - 1} = 0$; if $c_{v, s}^{l - 1} = \lambda_s^{l}$, the resource is at the margin.

In the optimal offering problem, since $\lambda_s^{l}$ is price-taking, resource $v$'s profit depends only on dispatch quantity, which is determined by comparing reported cost to $\lambda_s^{l}$.
Misreporting costs leads to either unprofitable dispatch (when reporting $\hat{c}_{v, s}^{l - 1} < c_{v, s}^{l - 1}$ and $\hat{c}_{v, s}^{l - 1} < \lambda_s^{l} < c_{v, s}^{l - 1}$) or missed profitable opportunities (when reporting $\hat{c}_{v, s}^{l - 1} > c_{v, s}^{l - 1}$ and $c_{v, s}^{l - 1} < \lambda_s^{l} < \hat{c}_{v, s}^{l - 1}$).
Therefore, truthful cost reporting maximizes profit, establishing IC at level $L$, consistent with traditional pool-based electricity markets.
\end{proof}

\section{Proof of Proposition 2 and Proposition 3}\label{app_proof_of_proposition_2}

\begin{proof}
For the same special case as in Proposition 1, the KKT conditions at level $l$ include:
\begin{equation}\label{eq:l_stationary_condition_vpp_v}
  \frac{\partial L}{\partial x_{v, s}^{l - 1}} = c_{v, s}^{l - 1} - \lambda_s^{l} - \underline{\mu}_{v, s}^{l - 1} + \overline{\mu}_{v, s}^{l - 1} = 0, \quad \forall v, s
\end{equation}\vspace{-2ex}
\begin{equation}\label{eq:l_stationary_condition_vpp_s}
  \frac{\partial L}{\partial x_{s}^{l}} = - \lambda_s^{l + 1} + \lambda_s^{l} = 0, \quad \forall s
\end{equation}
and complementary slackness conditions similar to (\ref{eq:complementary_slackness_condition}).

From (\ref{eq:l_stationary_condition_vpp_s}), we have $\lambda_s^{l} = \lambda_s^{l + 1}$ for all $s$, establishing Proposition 3.
If IC holds at level $l+1$, then $\lambda_s^{l+1}$ is not affected by the bid of individual resources, and by (\ref{eq:l_stationary_condition_vpp_s}), so is $\lambda_s^{l}$.
The same reasoning as in Proposition 1 applies: from (\ref{eq:l_stationary_condition_vpp_v}) and complementary slackness, the dispatch rules are identical, and misreporting costs leads to either losses or missed opportunities.
Therefore, truthful reporting is optimal, establishing IC at level $l$ given it holds at level $l+1$.
\end{proof}




\ifCLASSOPTIONcaptionsoff
  \newpage
\fi



\bibliographystyle{IEEEtran}
\bibliography{reference}
\end{document}